# On the possibility of electronic DNA nanobiochips.

V. D. Lakhno and V. B. Sultanov

*Institute of Mathematical Problems of Biology, Russian Academy of Sciences,*

*142290, Pushchino, Moscow Region, Russia.*

**Abstract**

We have considered a theoretical possibility for the development of a nanobiochip the operation principle of which is based on measuring conductance in single-stranded and double-stranded DNA. Calculations have demonstrated that in the majority of cases the conductance of double-stranded nucleotides considerably exceeds that of single-stranded ones. The results obtained are in agreement with recent experiments on measuring the oligonucleotide conductance.

It has been shown that an electronic biochip containing eleven nucleotide pairs will recognize $\approx 97\%$ sequences. It has also been demonstrated that the percentage of identifiable sequences will grow with the sequence length.

DNA microarrays, also known as biochips and genechips, have become the latest and most efficient method to analyze thousands of genes simultaneously. Potential applications are countless, including gene expression analysis, studies of genetic disorders, drug development, even the solution of mathematical problems[1-2]. Development of the method is concerned with further miniaturization, i. e. increase in the number of spots on the biochip. Progress in this field goes on at about the same rate as which miniaturization in computer technology advances (Moore's law).

The methods for processing information readout from biochips may come up against the limits of miniaturization still earlier than the computer technology (where the circuits of 0.1 $\mu$m size are presently used) will do, since readings from biochips are currently taken by measuring the intensity of a signal from an individual spot and analyzing changes in its intensity or color. This places a fundamental limitation on the accuracy of the measurements which is determined by the wavelength of the light emitted by a fluorescent marker used in the biochip technology, i. e. $\cong 1$ $\mu$m. At present, technically attainable size of a spot is about 20 $\mu$m.

One way to overcome this limitation is to develop electronic biochips from which one could read out detailed information by measuring the conductance of an individual fragment of a DNA molecule used in biochips. This would enable the biochip technology to break through into the nanometer region since in this case the resolution of the readings would be determined by the characteristic size comparable with the diameter of a DNA molecule, i. e. . $\cong 2$ nm.

The aim of this work is to study the possibility for the development of a nanobiochip, the operation principle of which is based on measuring the DNA conductance. Theoretically, a nanobiochip based on measuring the conductance will be efficient if the conductance of a single-stranded DNA will differ considerably from that of a DNA duplex obtained as a result of hybridization of the single-stranded DNA with a complementary strand. The paper just deals with elucidation of this problem.

Presently there are a great many approaches to development of biosensors which could discriminate between a single-stranded DNA and its hybrid with a complementary strand. These approaches include optical[3], electrochemical[4-9] and microgravimetrical[10] methods. Without going into details we will just notice that all the approaches have their advantages and drawbacks. The chief drawback of the above methods is that they all require rather a great number of similar sequences on a chip which presents an obstacle for substantial miniaturization.

The model of a nanobiochip considered here is a single-stranded DNA fragment whose ends are attached to electrodes as shown in Fig. 1 a). Fig. 1 b) presents a nucleotide sequence Fig. 1 a) in hybrid with a complementary strand in which charge transfer is determined by matrix elements of the transitions between neighboring nucleotides. The values of matrix elements in the case when charge is carried by holes are given in Table 1.

Our scheme of a nanobiochip where the ends of only one DNA strand are fixed reflects an experimental situation[14] (and references therein). In these works the ends of only the DNA strand attached to electrodes are modified (by thiol group).

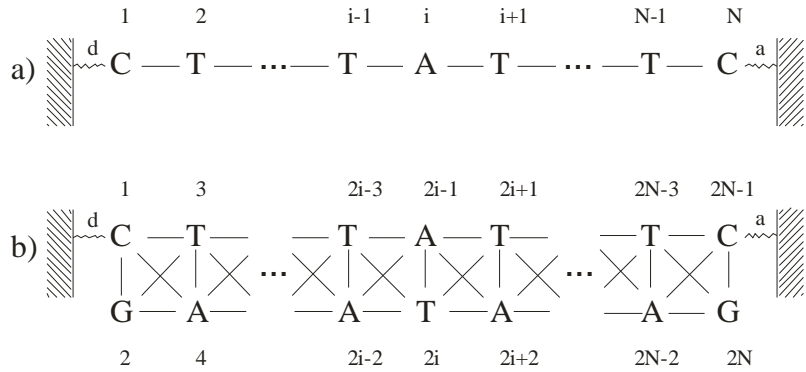

Fig. 1. a) – single-stranded DNA; b) – DNA duplex with all nearest interactions.

The charge transfer can be calculated with an effective tight-binding Hamiltonian $\hat{H}$ with diagonal elements determined by oxidation energies of the bases and off-diagonal elements equal to the corresponding charge coupling of neighboring bases[15]:

$$\hat{H} = \hat{H}^0 + \hat{V},$$

$$\hat{H}^0 = E_d |d\rangle\langle d| + \sum_{i=1}^{2N} E_i |i\rangle\langle i| + E_a |a\rangle\langle a|,$$

$$\hat{V} = v_{d,1} |d\rangle\langle 1| + v_{a,2N-1} |2N-1\rangle\langle a| + \hat{V}_M,$$

$$\hat{V}_M = \sum_{i \neq j} v_{ij} |i\rangle\langle j|,$$

where $V_{d,1}$, $V_{a,2N-1}$ are the coupling matrix elements between donor (d) and nucleotide (1) and acceptor (a) and nucleotide (2N-1) respectively; $\hat{V}_M$ is the bridge DNA Hamiltonian, where $v_{ij}$ are the matrix elements between neighboring nucleotides; $E_d$, $E_a$ and $E_i$ are hole energies on donor, acceptor and i-th nucleotide respectively. This Hamiltonian takes into account all the links between neighboring nucleotides (including off-diagonal ones). The lack of some off-diagonal links in Table 1 stems from the fact that their value is small and they cannot be found by quantum-chemical calculations with good accuracy.

Table 1[*]. Matrix elements of an electron transition (in eV) between neighboring nucleotides in DNA – duplexes[11-12].

Transitions inside the strand.

| 5' → 3' | A | C | T | G | | 3' → 5' | A | C | T | G |
|---|---|---|---|---|---|---|---|---|---|---|
| A | 0.030 | 0.061 | 0.105 | 0.049 | A | | 0.030 | 0.029 | 0.086 | 0.089 |
| C | 0.029 | 0.041 | 0.1 | 0.042 | C | | 0.061 | 0.041 | 0.076 | 0.110 |
| T | 0.086 | 0.076 | 0.158 | 0.085 | T | | 0.105 | 0.1 | 0.158 | 0.137 |
| G | 0.089 | 0.110 | 0.137 | 0.084 | G | | 0.049 | 0.042 | 0.085 | 0.084 |

Transitions between the strands.

| 5' → 5' | A | C | T | G | | 3' → 3' | A | C | T | G |
|---|---|---|---|---|---|---|---|---|---|---|
| A | 0.035 | 0 | 0.016 | 0.021 | A | | 0.062 | 0 | 0.016 | 0.021 |
| C | 0 | 0.0007 | 0 | 0 | C | | 0 | 0 | 0 | 0 |
| T | 0.016 | 0 | 0.002 | 0 | T | | 0.007 | 0 | 0.002 | 0 |
| G | 0.021 | 0 | 0.009 | 0.019 | G | | 0.021 | 0 | 0 | 0.043 |

Oxidation potentials[13] (in acetonitrile solution).

| A | C | T | G |
|---|---|---|---|
| 1.69 | 1.9 | 1.9 | 1.24 |

Transitions inside the site.

| C – G | A – T |
|---|---|
| 0.050 | 0.034 |

[*] Table 1 was reproduced from work[16].

To compare the conductances of the single-stranded DNA fragment and the DNA duplex presented in Fig. 1 let us consider their ratio f:

$$f = \frac{R_{1N}^2}{R_{1,2N-1}^2}, \text{ where}$$

$$\hat{R} = (E - \hat{H})^{-1}$$

is a resolvent, corresponding to appropriate $\hat{H}$, $R_{ij}$ are appropriate Green's functions[15], E is a complex variable, the real part of which is interpreted as a hole energy; the parameters of electrodes playing the roles of a donor and an acceptor are taken to be: $E_a = E_d = E_G$, $v_{d1} = v_{G1}$, $v_{a,2N-1} = v_{G,2N-1}$. In other words we consider the situation when the energy parameters of electrodes are close to corresponding values of guanines. Notice, that the longer is the sequence used in a biochip, the less important is a particular choice of the donor and acceptor parameters.

The ratio f, which is a function of the sequence, was calculated for all possible nucleotide sequences of length N = 8, 9, 10, 11. The number of such sequences is $4^N$. The function f was calculated for E=-0.5 eV, which was found[16] from the condition of the best agreement of the calculated charge transfer rates with the experimentally measured ones for some oligonucleotide sequences. The sequences in which f differed from 1 by less than 10% : |f-1|<0.05 (in the absence of interactions between the strands f≡1) were considered to be indistinguishable. The results of the calculations are presented in Table 2. The second column of Table 2 shows that in the majority of cases the conductance of double-stranded nucleotides considerably exceeds that of single-stranded ones. This conclusion is confirmed by experimental data[14].

| Table 2. The results of the calculations. | | | | |
|---|---|---|---|---|
| N | percent f<1 | percent of indistinguishable | max(f), sequence | min(f), sequence |
| 8 | 75.5% | 7% | $1.3 \times 10^6$, CAAATGTG | $3.5 \times 10^{-7}$, CCCCCCCG |
| 9 | 79.9% | 5.1% | $7.6 \times 10^7$, ACTTGAACC | $1.5 \times 10^{-8}$, CCCCCCCG |
| 10 | 82.4% | 3.8% | $1.4 \times 10^{10}$, CACGCAGGAG | $6 \times 10^{-10}$, CCCCCCCCG |
| 11 | 85% | 2.8% | $4.2 \times 10^{13}$, TGCGCCTTCCC | $2.5 \times 10^{-11}$, CCCCCCCCCCG |

The third column of Table 2 shows that the percentage of indistinguishable sequences decreases as the sequence length N grows. The results obtained testify to the perspectiveness of the development of nanobiochips whose operation principle is based on the conductance measurement. The accuracy of operation of such biochips will be the higher, the larger will be the sequence length.

As was shown above, the charge transfer rate in DNA molecule strongly depends on the type of a nucleotide sequence. However, current theoretical estimations of charge tunneling probabilities in specific double helixes of olygonucleotides made with the use of matrix elements of overlap integral of electronic orbitals of neighboring nucleotides in DNA are based only on a significantly idealized (nearly to planar) model of base – pairing. They do not take into account either the significant heterogeneity of nucleotide geometry parameters in a DNA molecule observed in X-ray experiments or the existence of large "propeller" and "buckle" deformations of the planar structure of H-bonding bases. These deformations, as is known, can reach $50°$[17-20]. In addition DNA thermal fluctuations break the global helix symmetry and may localize the wavefunction (HOMO and/or LOMO) to different areas as a function of time. It was shown[21-23] that correct description of the formation of a DNA double helix requires consideration of intrinsic polymorphism of H-bonding of canonical base pairs which arises from a bistable, pyramidal structure of amino-substituted nitrous bases. Calculation of olygonucleotide double strands in the gas phase and in pure water solution have demonstrated the existence of rather a large "propeller-like" and "step-like" deformations of H-bonding base pairs in double helix stacking conditions. It is this initial non-planar structure of canonical DNA base pairs that initiates their non-coplanar packing in the structure of double helix and, hence, is responsible for the dependence of the form of a DNA helix on the nucleotide sequence. Based on these results we can conclude that probabilities of charge transfer in DNA with due regard for real distortions of base pairing in the structure of double helixes will greatly differ from those obtained in studies with the idealized model of planar base pairs.

In conclusion we note that the parameters (matrix elements) used by us approximate and point out some physical factors which also can lead to their deviation from our parameter values (i. e. with a specific substrate, environment, etc.). In a real nanobiochip the complementary DNA strand can come in contact with electrodes. To rule out this possibility one should take measurements with several identical chips in parallel. These difficulties can be overcome in calculations of a specific laboratory nanobiochip. Such studies which we are planning to carry out in future will more correctly describe the dependence of the electron tunneling rate in a DNA molecular with a nucleotide sequence under study.